\documentstyle[aps,multicol,epsfig,eqsecnum]{revtex}
\begin{document}
\draft

\title{Sharp gene pool transition in a population affected by phenotype-based 
selective hunting}

\author{E. Brigatti $^{\dag,}\thanks{E-mail address: edgardo@cbpf.br}$,
         J.S. S\'a Martins $^{\star}$,   
         and I. Roditi $^{\dag}$ 
        }  
\address{$\dag$Centro Brasileiro de Pesquisas F\'{\i}sicas, Rua Dr. Xavier 
  Sigaud 150, 22290-180, Rio de Janeiro, RJ, Brasil} 
\address{$\star$Instituto de F\'{\i}sica, Universidade Federal Fluminense, 
  Campus da Praia Vermelha, 24210-340, Niter\'oi, RJ, Brasil}
  
 \maketitle
 \widetext

\begin{abstract}
We use a microscopic model of population dynamics, a modified version 
of the well known Penna model, to study some aspects of microevolution. This 
research is motivated by recent reports on the effect of selective hunting on the 
gene pool of bighorn sheep living in the Ram Mountain region, in Canada. Our 
model finds a sharp transition in the structure of the gene pool as some 
threshold for the number of animals hunted is reached.
\end{abstract}  
  

\begin{multicols}{2}

\section{Introduction}

The impact of anthropic activities on ecologic equilibria is a common and urgent 
problem that challenges the scientific community to find new and sharper solutions 
for sustainable resource management. In particular, new data and studies have 
pointed out how the traditional approaches, focused only on demographic and 
ecological factors, ignoring the possibility of evolutionary changes, are 
hiding part of the problem \cite{Ashley}.
A clear evidence for such phenomena is particularly well demonstrated by the 
effects of overfishing \cite{Law}. Actually, 
we can look at this activity as a good paradigm, as some sort of large scale 
experiment on life history evolution, that gives a strong support to the thesis of 
the speeding up of evolutionary processes as a result of human activities. In 
fact, fishing causes a highly selective mortality depending on a particular trait 
(size). The variability of a trait that confers a difference in survival and, as 
a consequence, in reproductive success, added to the fact that this trait 
variation is inherited, causes the population to evolve in a direction that 
opposes the trait bias of harvesting \cite{Conover}. We can easily find other 
examples in many different areas of biology. It is well known, for instance, that 
bacteria are able to rapidly develop resistance to many antibiotics. Evolution of 
immunity to such a drug, that could occur, albeit with a very low probability, 
even without human intervention, can be reached within a few years of the 
commercial usage of a new antibiotic \cite{Palumbi}. By the same token, the 
attempt to control the overpopulation of European domestic rabbits in Australia, 
that grew up to a plague, through the introduction of myxomatosis was frustrated 
by the effect of a fast evolutionary response to the disease.
Even if there is no clear quantitative measure of the speed with which evolution  
responds, it is possible to find growing evidence that evolutionary shifts are 
sometimes very fast, comparable to the lifespan of a human being (microevolution). 
Support to this consideration is given by recently published data related to a 
long term study of the effects of trophy hunting on a population of bighorn sheep 
on the Ram Mountain, Canada \cite{Coltman1}. From this study it emerges that, in a 
timespan of just 30 years, selective hunting based on some phenotype character - 
horn size in this particular case - can have caused the depletion of the genes 
that confer rapid body and horn growth. In fact, an observed decline in mean 
breeding values for weight and horn size indicates a microevolutionary response to 
hunting selection.

\begin{figure}[p]
\begin{center}
\vspace*{0.8cm}
\resizebox{0.4\textwidth}{!}{\includegraphics{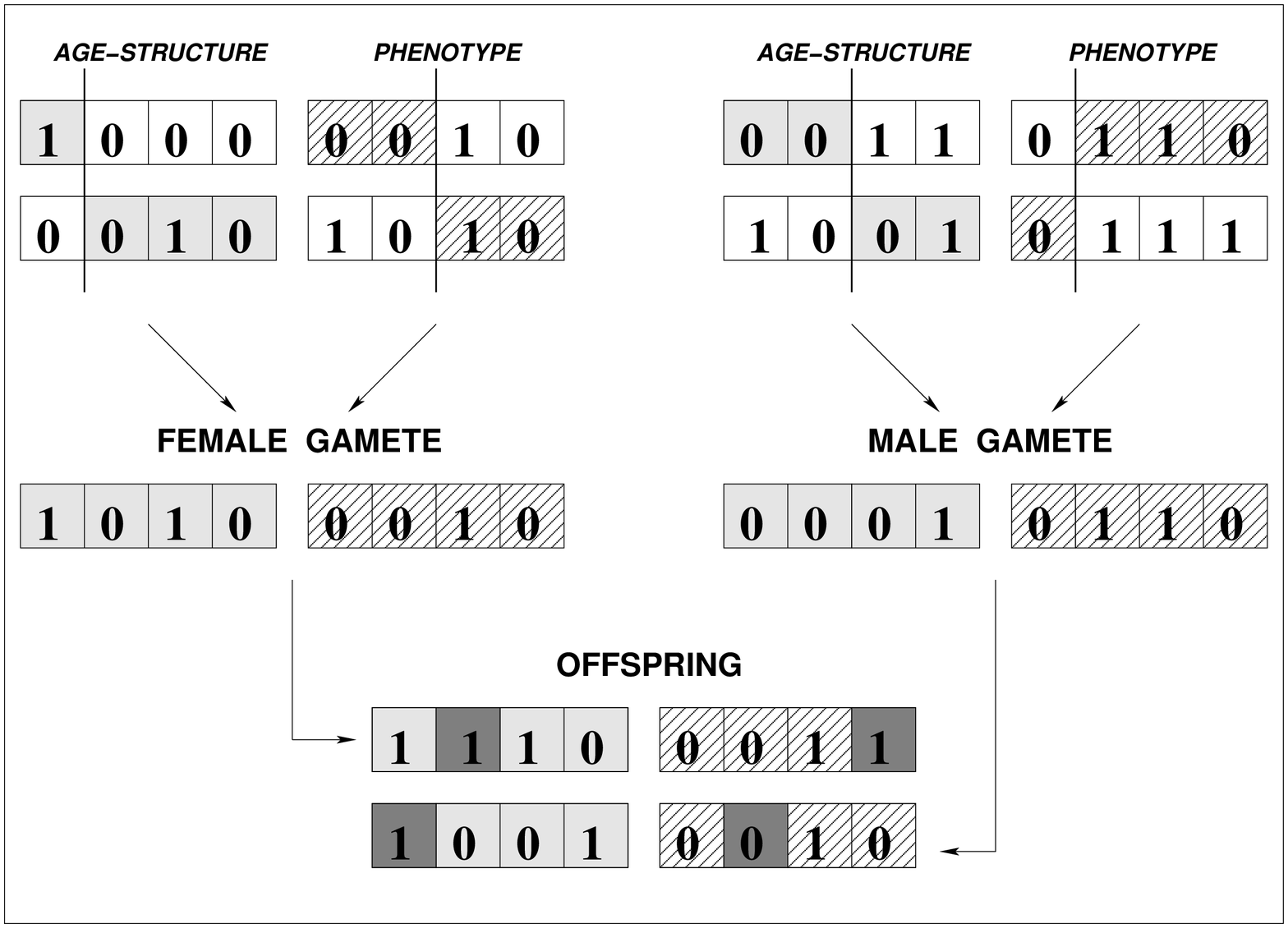}}
\vspace*{0.4cm}
\end{center}

\caption{\small Representation of the reproductive cycle. 
The diploid genome is present with 
its age-structured part (light-shaded background) 
and the bit-strings that encode 
for phenotype (diagonal stripes background).   
In the first line the crossing-over is performed.
 In the second passage the haploid gamete 
is chosen.
Finally, some new mutation were added 
(dark-shaded squares) and the gametes 
combine to form a new genetic strand.
}
\label{fig_Exa}
\end{figure}

The main aim of this work is to show that a simple computational model of 
population dynamics, the Penna Model \cite{Penna}, is capable of reproducing such 
process. To be specific, our interest is to show how can we obtain a stable 
solution for a population subjected to microevolution and to give a full 
description of its dynamics.
For this purpose, we will describe the dynamics of a diploid sexual population, 
representing each individual by its diploid genome, simulated by two sets of 
coupled bit-strings. An allele is encoded by two homologous bits in each of these 
bit-strings. The collective behaviour of these agents is dictated by a set of 
simple rules that we sum up in what follows. Introduction of new individuals in 
the population is obtained through a reproductive cycle that simulates birth. 
Each individual of a reproductively active couple generates, through a meiotic 
cycle with crossing-over two haploid gametes that, after the introduction of some 
mutations, combine to form a new genetic strand.
Crossing is obtained cutting at a random position each string of the parent's 
genome and combining the left part of one with the right part of the other. In so 
doing, two new combination of the original genes are generated, and the haploid 
gamete is chosen randomly to be one of these  (see Fig. \ref{fig_Exa} ).

The passage of time and the death of agents follow rules inspired by biological 
ageing as described by Medawar's hypothesis of 
accumulation of bad mutations. A position (locus) of the chronological 
(age-structured) piece of the genome is read at each time step. If an active 
mutation is found at this locus, it is added to the current number of harmful 
mutations; the individual dies when this amount reaches some pre-determined 
threshold value. In order to be active, a harmful mutation must occur at a 
homozygote locus, or at a heterozygote one for which the harmful allele is 
dominant. The number of loci where the allele represented by a bit set to $1$ is 
dominant is fixed, and their position in the genome string is chosen randomly at 
the beginning of the simulation.

The biological and computational necessity of studying finite populations imposes 
the introduction of another death factor. It is usually represented by a 
density-dependent mean-field death probability, called Verhulst factor, 
proportional to the current size of the population.  

The full simulated genome can include, other than the age-structured bit-strings 
used to introduce the biological clock of the individuals, other pairs of 
bit-strings that encode for phenotype traits that are responsible for intraspecific 
and/or environmental interactions. With this technique it is possible to establish 
a computational representation of competition and/or sexual selection 
\cite{Pheno,Ticona,Karen}. In our present study, we will use just one extra 
bit-string to represent a single biological trait. A phenotype value is attributed to 
it by summing over all the active mutations present in the bit-string that encodes 
this trait - again, we consider as active a mutation at a homozygous locus, or at a 
heterozygous one where the mutated allele is dominant. 

According to this procedure, the phenotype value is an integer between $0$ and 
$32$, and is different from the simple sum of all the ones present in the bit-string 
that determines the genetic distribution. We will refer to this latter number as 
the trait bit-string value. The essential difference lies in that the phenotype 
value takes into account the dyploid nature of the genome, through the effects of 
dominance and homozygose. On the opposite, the trait bit-string value is simply 
related to the frequency of the $1$ allele in the genome.

Bits of the trait bit-strings can mutate from $0$ to $1$ or from $1$ to $0$, as 
opposed to ageing bit-string, that can undergo only bad mutations.

\section{Model and methods}

In a previous publication \cite{Penna2}, the simplest implementation of the Penna 
model was able to make predictions relative to changes caused by overfishing on a 
population with a strong relation between age and size. This fact allowed the 
usage of the age-structured bit-strings of the genome alone for a full description 
of the relevant interactions.

The idea of the present work is to build up a simple toy model, inspired by the 
gene pool dynamics of the Ram Mountain population, capable of representing just 
the fundamental features of a real population. 

For this reason, we have chosen to represent each individual by a genome with just 
two pieces: the first is the age-structured bit-string, while the second bit-string 
determines the individual phenotype. The classical rules of the Penna model, with a 
standard logistic Verhulst factor \cite{Pheno}, are applied to this population. 
Under these conditions, the population dynamics obtained is simple and well known: 
the age-structured part of the genome causes the population to age in accordance 
with Gompertz law. On the other hand, the trait bit-string value, that does not 
feel the effect of any interaction, reaches a Gaussian shaped distribution with 
$16$ as the mean value. 

We now add to this simple basic model the two main forces that drive the dynamics 
of the phenotype distribution: sexual selection correlated with the phenotypic 
aspect and a phenotype-based selective hunting.

Our simulations are inspired by the observational data related to the population 
of the Ram Mountain \cite{Coltman2}, where sexual selection in bighorn rams 
operates in such a way that mating success increases with age, horn length and 
body size, tending to concentrate the paternities in those more favored rams. In 
our simulations we represent all the phenotypical dependence of mating success 
by one single trait. For simplicity, mating success has no age dependence in our 
implementation.

\begin{figure}[p]
\begin{center}
\vspace*{0.8cm}
\resizebox{0.4\textwidth}{!}{\includegraphics{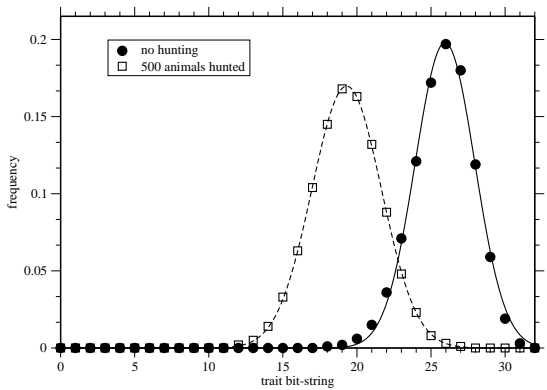}}
\vspace*{0.4cm}
\end{center}

\caption{\small Trait bit-string distribution for the two equilibrium states (no 
hunting and hunting pressure equal to $500$). The data are well fitted by 
Gaussians with mean values $25.9$ and $19.3$, and standard deviations $2$ and 
$2.4$ respectively. The parameters used in the simulation are: the Verhulst 
parameter ($130000$), the initial population ($2000$), the minimum reproduction 
age ($8$), the maximum reproduction age ($32$), the number of offsprings ($1$), 
the threshold value ($3$), the same number of mutations ($1$) and dominant loci 
($6$) for the two bit-strings. We have averaged over $10$ different realizations in 
each case.}
\label{fig_Dis}
\end{figure}

The first ingredient that is essential to affect the general distribution of the 
phenotype is to allow for paternity concentration. For this reason it is important 
to allow a male to mate more than once in each reproductive cycle. At each time 
step, each female that is reproductively active makes a non-random choice of her 
mating partner. The selection is made by choosing among $20$ males the one that 
has the highest phenotype value (extreme dynamics). This sexual selection causes a 
drift of the trait bit-string distribution towards higher values. In figure 
\ref{fig_Dis} we can see that the distribution, after 100000 time steps, 
is centered around $26$. 

The hunting pressure we simulate was inspired by the conditions and laws 
prescribed in the Ram mountain region \cite{Coltman1}. These state that it is 
allowed to hunt only rams older than four years and ``full-curl'' trophy rams, up 
to a maximum number of animals. In our simulations, we hunt male individuals 
older than $4$ and with a phenotype value bigger than 16. The harvesting is 
implemented in the following way: we do a number of random attempts 
(approximately equal to the population dimension) of finding individuals that 
satisfy such characteristics and we stop the hunting when the established 
threshold number of animals killed is reached.

With this simulation setting, we leave the population evolving, with no harvesting 
selection, for 50000 time steps. After this first equilibrium is reached, we 
switch on the hunting selection. We estimate that after 50000 time steps further a 
final equilibrium is obtained.

In the simulations, particular attention must be payed to the population size. The 
real biological population of the Ram Mountain, during the years of observation, 
was approximately $140$ animals, with rather large fluctuations over the years. 
The simulation of such a small population is really problematic. Not only is it 
difficult to do a confident statistical analysis of the results, but also the 
size of the fluctuations destroys the key elements of the model. The general 
dynamic, in fact, is no longer driven by the bit-string dynamics, but by the 
Verhulst factor instead,  and it is also difficult to find a phenotype 
equilibrium distribution. For all these reasons we decided to describe the 
behaviour of a population of about 10000 individuals, considering this the 
smallest population that it is still suitable for a study with this model and 
that can allow for a consistent statistical approach. We have performed 
simulations with about 250000 individuals but it was not possible to point out 
differences or possibles finite scale effects.

\section{Results}

We say that the population has undergone a significant variation in its gene pool 
when the corresponding mean value of the trait bit-string distribution has changed 
by more than the value of its standard deviation. By this token, the mean value of 
the trait bit-string distribution is used to define the state of the system. 

We have decided to use this parameter instead of the phenotype value to keep our 
focus on evolutionary changes and variation at the genome level. The choice is, as 
a matter of fact, immaterial, and the results are qualitatively identical because, 
in our model, the relation between phenotype and genome is over-simplified and 
direct, and does not take into account environmental factors.

As we can see in Fig. \ref{fig_Dis} the fundamental state of the system is a 
distribution centered in $26$ with standard deviation of $2$. If the hunting 
selection is switched on, a drift towards lower values is caused. In particular, 
if we select a hunting threshold of $230$ animals for each time step, the 
distribution has its lowest mean value ($19$). From this point on, increasing the 
number of animals hunted will not cause any changes in the distribution. The 
distribution with mean $19$ is thus some sort of equilibrium, or statistically 
stationary state, and limitations imposed on the hunting process will not affect 
any further the gene pool of the population.

\begin{figure}[p]
\begin{center}
\vspace*{0.8cm}
\resizebox{0.4\textwidth}{!}{\includegraphics{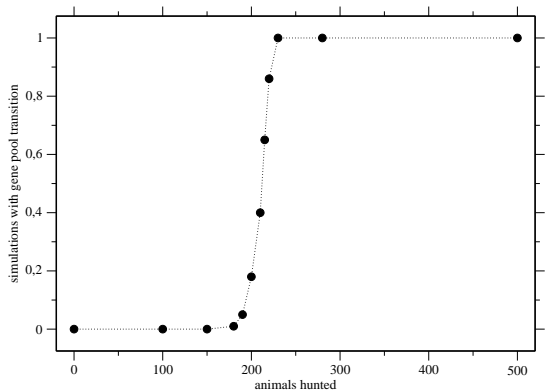}}
\vspace*{0.4cm}
\end{center}

\caption{\small  Fraction of simulations that undergo a transition as a function 
of the number of animals hunted.}
\label{fig_Ratio}
\end{figure}

We performed a hundred different simulations, for each set of parameters, to 
investigate the behaviour of the gene pool as a function of the number of animals 
hunted. The results are shown in Figure \ref{fig_Ratio}. It emerges that, for 
values smaller than $200$, the fraction of simulations with a mean value of the 
trait bit-string distribution smaller than $23$ is negligible. For $200$ kills, 
in $20\%$ of the simulations the model suffers a transition of the trait 
bit-string distribution to reach a mean value close to $19$. By increasing the 
number of animals killed, the number of simulations that undergo the transition 
also increases, and, for values larger than $230$, all the simulations end up 
with a distribution with a mean value close to $19$. An interesting and not so 
intuitive result is that not all the possible states are visited by the phenotype 
distribution as the number of animals hunted is varied. There is a forbidden 
region between $21$ and $23$, never visited by the mean value of the trait 
bit-string distribution. As a result, the model has only two different stationary 
states. 

\begin{figure}[p]
\begin{center}
\vspace*{0.8cm}
\resizebox{0.4\textwidth}{!}{\includegraphics{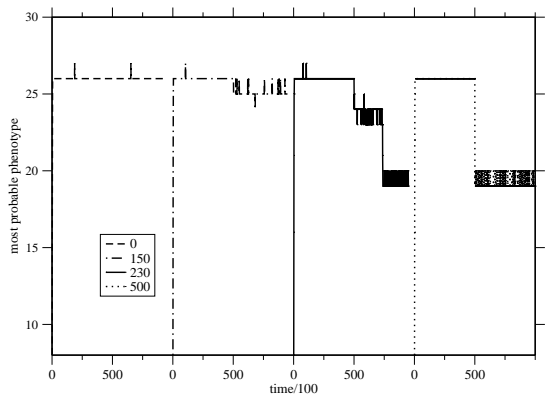}}
\vspace*{0.4cm}
\end{center}

\caption{\small  The most probable value (mode) assumed by the trait bit-string 
distribution as a function of time. When there is no hunting it reaches the value 
$26$ and does not change any more. For $150$ animals hunted, from time step 
$50.000$ on, this value moves to $25$. For $230$, it changes to $24$ before 
stabilizing at the state of equilibrium at $19$. In the last simulation, for $500$ 
animals hunted, the mode undergoes a sharp transition towards $19$. The time 
coordinate is shifted to a different origin for each simulation for clarity.}
\label{fig_Time}
\end{figure}

If we look at the dynamical behaviour of the model (see Figure \ref{fig_Time}),
it is possible to notice that the evolution of the system is really abrupt, and 
only $100$ time steps are sufficient to drive the trait bit-string values from one 
state to the other (for $500$ animals hunted, all the simulations reached the 
transition in less than $100$ time steps). This fact is particularly interesting 
because it shows how fast the sharp transition that we are describing occurs. 
For this reason, we claim that our model can well represent the phenomenon of 
microevolution that has motivated our work. A fast transition is not the only 
possible outcome though. In fact, in some simulations where the number of animals 
hunted is between $210$ and $230$, the trait bit-string value becomes metastable 
before reaching the stationary state, and the time-scale of the transition becomes 
large.
 
\section{Conclusion}

In this paper, we provide an example of how studies of microevolutionary 
processes may be undertaken with the usage of microscopic models of population 
dynamics. By a careful selection of the key ingredients, we show that a toy model 
that mimics the main features of some particular gene pool dynamical behaviour 
can single out the dominant structures of its trajectories in phase space. In our 
particular case, these features relate to a biological trait that identifies 
preferred mating partners and also hunting trophies. The competition that results 
from this situation generates a transition in the gene pool repertoire of the 
population as some threshold number of animals are killed, both in the results of 
our simulations and in the observational data.

\section*{Acknowledgments} 

We thank the Brazilian agencies CAPES, CNPq, and 
grants from PRONEX (PRONEX-CNPq-FAPERJ/171.168-2003 and PRONEX-FAPERJ E-26/171/2003 ) and FAPERJ 
(E-26/170.699/2004) for partial financial support.

\end{multicols}

\end{document}